\documentclass[aps,prd,superscriptaddress,floatfix,nofootinbib,showpacs,amsmath,amssymb]{revtex4-1}
\usepackage[colorlinks=true,pdfstartview=FitV,linkcolor=blue,citecolor=blue,urlcolor=blue]{hyperref}

\usepackage[inline]{enumitem}
\usepackage[multidot]{grffile}  % allow the name of figures to include dots
\usepackage{dcolumn}
\usepackage{bm}
\usepackage{amsmath}
\usepackage{amsfonts}
\usepackage{amssymb}
\usepackage{color}
\usepackage{latexsym}
\usepackage{slashed} % slash mark
\usepackage{pstricks}
\usepackage{indentfirst}
\usepackage{mathrsfs}
\usepackage{multirow}
\usepackage{epsfig,psfrag}
\usepackage{subfigure}
\usepackage{mathtools}
\usepackage{setspace} % spacing
\usepackage[utf8]{inputenc} % accept utf-8 input encoding
\usepackage[scientific-notation=true]{siunitx} % comprehensive units
\graphicspath{{fig/}}

\newcommand{\overbar}[1]{\mkern 1.5mu\overline{\mkern-1.5mu#1\mkern-1.5mu}\mkern 1.5mu}

\makeatother

\allowdisplaybreaks% allow eqnarray breaks
\begin{document}
\title{Derivative Portal Dark Matter}
\author{Yu-Pan Zeng}
\email[E-mail: ]{zengyp8@mail2.sysu.edu.cn}
\affiliation{School of Electronics and Information Engineering, Guangdong Ocean University, Zhanjiang 524088, China}
\affiliation{School of Physics, Sun Yat-sen University, Guangzhou 510275, China}
    \begin{abstract}
	We propose a new kind of Dark Matter: Derivative Portal Dark Matter. This kind of Dark Matter connects to the Standard Model through a massive mediator, which links to the Standard Model in derivative form. The derivative of a mediator in momentum space corresponds to the mediated momentum, which vanishes in the zero momentum transfer limit. As a result, this kind of Dark Matter can evade stringent constraint from the Dark Matter direct detection while fitting the Dark Matter relic density observation naturally. We explore several UV complete models of this kind of Dark matter. What's more, we show that these models also survive from Dark Matter indirect detection and collider search.
    \end{abstract}
%\pacs{13.66.Jn, 95.35.+d} \maketitle %\tableofcontents
%\clearpage
\maketitle

\section{Introduction}
The ever-improving sensitivities of Dark Matter (DM) direct detection experiments have put the famous Weakly Interacting Massive Particles (WIMPs) DM models under pressure. Cold massive DM can explain the observed DM relic density through thermal production, with the requirement of weak interaction, which can be naturally interpreted as electroweak interaction possessed by the Standard Model (SM). This scenario also predicts DM with hundreds of ~$\mathrm{GeV}$ mass which can be explored by DM direct detection experiment. However, with the ever-improving sensitivities of DM direct detection experiments, no DM has been found. A smaller interaction coupling can be adopted to explain the null result of the DM direct detection search, while a smaller interaction coupling will also result in insufficient DM thermal production. This raises a question: how can WIMPs explain the DM direct detection null result without diminishing of DM thermal production? One way out of this is to introduce a cancellation mechanism which works in DM direct detection only. In recent years, there have been studies exploring models where direct detection interaction is cancelled by two scalar mediators in the zero momentum transfer limit~\cite{Gross:2017dan,Alanne:2018zjm,Karamitros:2019ewv,Jiang:2019soj,Zeng:2021moz}. In our previous work~\cite{Cai:2021evx} we have generalized the cancellation mechanism from scalar mediators to vector mediators\footnote{In ref.~\cite{Liu:2017lpo}'s Appendix B, there are also discussions of cancellation between vector mediators.}. Inspired by the proof of the cancellation mechanism in ~\cite{Cai:2021evx}, we propose a new kind of DM: Derivative Portal Dark Matter (DPDM), which also possesses this cancellation mechanism. This innovative type of DM is capable of evading the rigorous constraints imposed by direct detection and naturally account for the observed DM relic density.

The cancellation occurs when the scattering between DM and the SM fermions, mediated by two mediators, cancels each other out, resulting in an amplitude proportional to the momentum transfer. Therefore, the amplitude vanishes in DM direct detection since we usually adopt the zero momentum transfer limit in direct detection, while the DM relic density is not diminished since the momentum transfer in the annihilation process surpasses two times the DM mass and thus can not be disregarded. The usual way to prove the cancellation mechanism is to directly calculate scattering amplitude, which will be proportional to the momentum transfer. Alternatively, we can also prove it by noting that the momentum transfer is equal to the momentum of the mediators, which is equal to the derivative of the mediators in momentum space. Therefore, we can denote the interactions in models with the cancellation mechanism in the form of the derivative of the mediators (i.e., in the form of kinetic mixing between mediators). This allows us to see the cancellation property immediately. Therefore, in this work, we propose the DPDM model where the DM and the SM fermions are linked by the kinetic mixing between the $Z$ boson and a massive dark vector boson\footnote{The $Z$ boson can be replaced by another massive neutral-charged gauge boson which couples to the SM fermions.}, and the kinetic mixing between the two bosons ensures the cancellation property. There are lots of works have studied models where interaction between DM and the SM fermions comes from kinetic mixing term~\cite{Holdom:1985ag,Babu:1997st,Feldman:2007wj,Cheung:2009qd,Hook:2010tw,Davoudiasl:2012ag,An:2014twa,Batell:2009di,Lao:2020inc,Rueter:2020qhf}. While the distinctive point in our construction is that the dominant kinetic mixing is between massive vector bosons and the kinetic mixing between the photon and the dark vector boson should be naturally negligible (e.g., the kinetic mixing between the photon and the dark vector boson comes from two-loop corrections). The reason why the photon should be out of the picture is that the propagator of the photon contains a momentum transfer $t$ in its denominator, which will cancel the momentum transfer in the numerator and thus ruin the cancellation. Note in our previous work~\cite{Cai:2021evx} we focused on the cancellation mechanism between vector bosons, while in this work we aim at proposing the new DPDM model, which opens a new portal to DM. Besides possessing the cancellation mechanism, the models studied in this work are new and the Lagrangians are constructed directly in the form of derivative, enabling one to see the cancellation directly. 

The innovative feature of the DPDM model is that though the DM-SM fermion interaction comes from the derivative term (such as kinetic mixing term), while the kinetic mixing term between the massless photon and the extra massive gauge boson should be at least originates from two-loop corrections. This feature ensures validation of the cancellation mechanism in the DM direct detection. Then it raises a theoretical question: how to obtain such feature from a UV complete model naturally. This is difficult because we want non-negligible kinetic mixing between two massive mediators and negligible kinetic mixing between photon and mediator which link to DM. There are many works which studied UV complete model for generating the kinetic mixing in loop level~\cite{Holdom:1985ag,Dienes:1996zr,Rueter:2019wdf,Gherghetta:2019coi,Rizzo:2024bhn,Nomura:2021tmi}, which can give very small kinetic mixing between photon and mediator which link to DM. However, we still need not that small kinetic mixing between massive mediator which link to SM fermion and massive mediator which link to DM. Fortunately, there are particles in the SM which link to the $Z$ boson rather than the photon, therefore we will use these particles to build UV complete DPDM models, which generate kinetic mixing between an extra gauge boson and the $Z$ boson (the photon) in loop level (two-loop level).

In Sec.~\ref{sec:DPDM} we propose the DPDM model and discuss its cancellation mechanism. In Sec.~\ref{app:uv} we explore the UV origination of the DPDM model and constructed three DPDM models. We explore phenomenology constraints on DM in Sec.~\ref{sec:phe} and conclude in Sec.~\ref{sec:con}. 

\section{Derivative Portal Dark Matter} \label{sec:DPDM}
\begin{figure}[ht]
    \centering
    \includegraphics[width=0.4\textwidth]{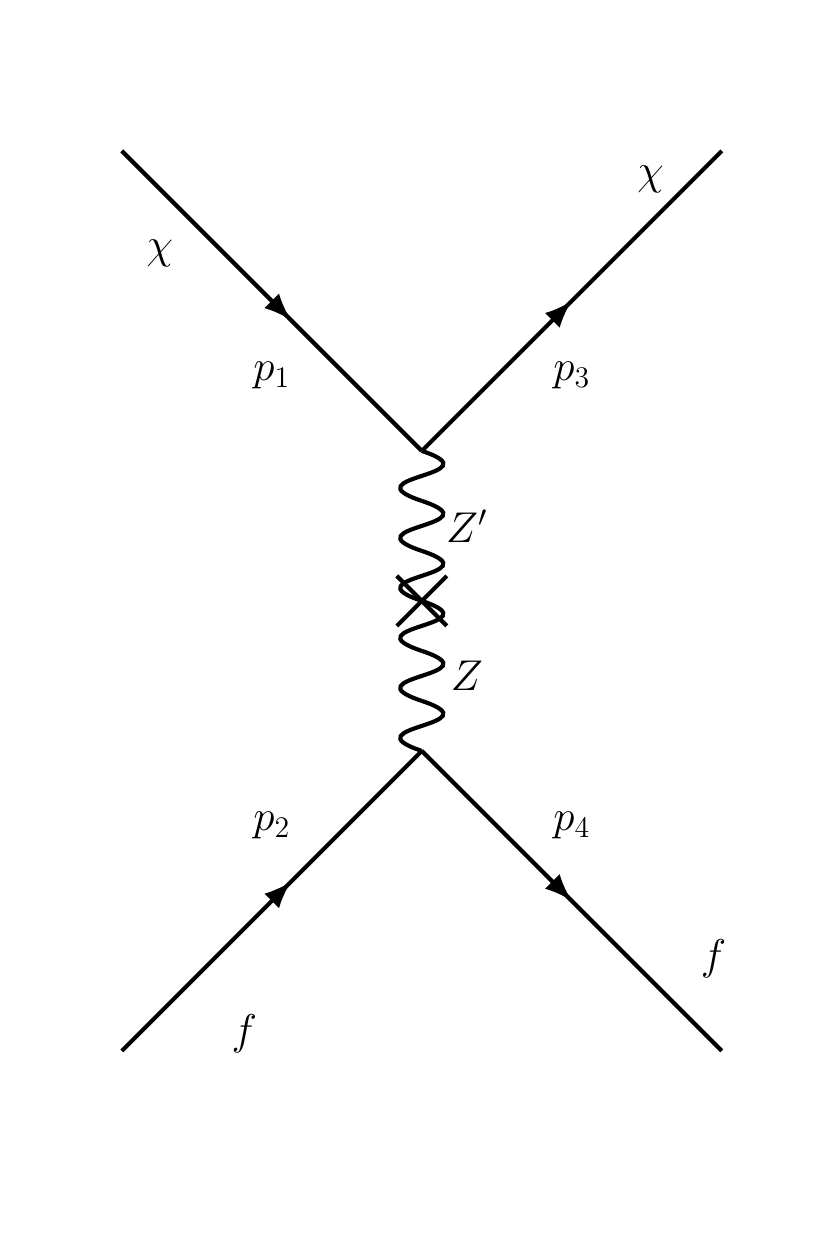}
    \caption{DM-SM fermion scattering in the DPDM model}
    \label{fig:DPDMfd}
\end{figure}

\begin{figure}[ht]
    \centering
    \includegraphics[width=0.5\textwidth]{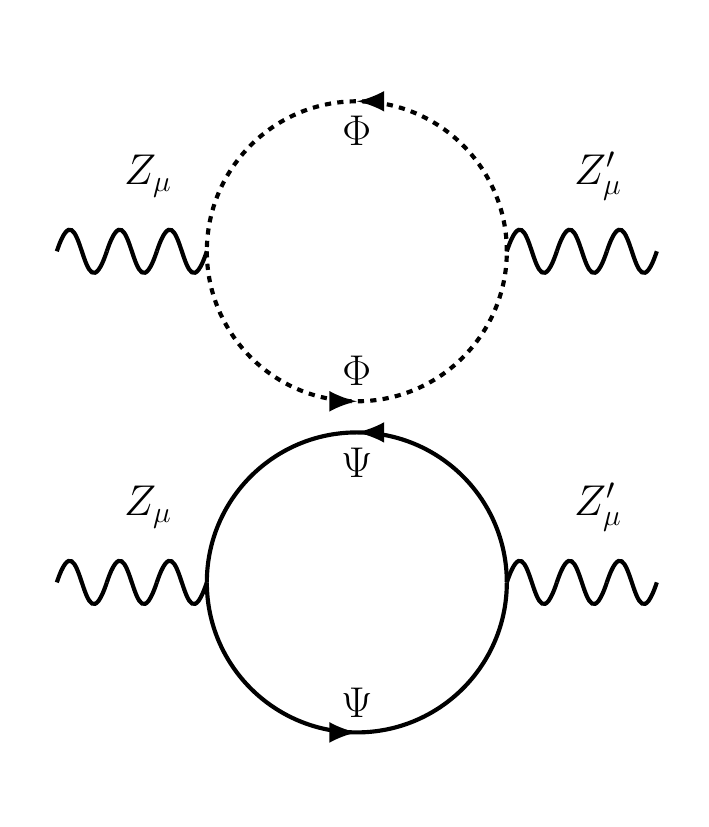}
    \caption{Derivative portal originating from loop corrections.}
    \label{fig:DPDMlo}
\end{figure}

The key Lagrangian of the DPDM model is
\begin{eqnarray}
    \mathcal{L}\supset J_{f}^{\mu}Z_{\mu}-\frac{\epsilon}{2} Z^{\mu\nu}Z_{\mu\nu}^{\prime}+J_{DM}^{\mu}Z^{\prime}_{\mu},
    \label{DPDM}
\end{eqnarray}
where $Z_{\mu}$ and $Z_{\mu}^{\prime}$ are massive vector mediators, while $J^{\mu}_{f}$ and $J_{DM}^{\mu}$ are the current of the SM fermions and DM respectively. $\frac{\epsilon}{2} Z^{\mu\nu}Z_{\mu\nu}^{\prime}$ is the derivative portal which connects the SM and the dark sector. Then the dark matter SM fermion scattering is depicted by Fig.~\ref{fig:DPDMfd},
where we use $\chi$ and $f$ to denote DM and the SM fermion respectively. Since the derivative portal contains derivative of mediators, which in momentum space is equal to the mediators' momentum, the scattering amplitude will be proportional to the mediators' momentum and thus the momentum transfer $t$:
\begin{eqnarray}
    i\mathcal{M}\propto \frac{p_1-p_3}{t-m_{Z^{\prime}}^2}\frac{p_4-p_2}{t-m_{Z}^2}=\frac{t}{(t-m_{Z^{\prime}}^2)(t-m_{Z}^2)},
    \label{amplitude}
\end{eqnarray}
where $m_{Z}$ and $m_{Z^{\prime}}$ represent the mass of $Z$ and $Z^{\prime}$ boson. Therefore the amplitude goes to zero in the zero momentum transfer limit\footnote{The usual way of proving the cancellation mechanism can be seen in Appendix~\ref{app:uproof}.}. From Eq.~\eqref{amplitude} we see that when the mass of one mediator goes to zero, there will be a $t$ in the denominator which will cancel the $t$ in the numerator and thus ruin the $t$-proportional property. Hence we will not want direct photon kinetic mixing in the derivative portal, and we will show that if the kinetic mixing between photon and the extra gauge boson which link to DM originating from two (or higher) loops level it will also preserve the $t$-proportional property. Now let us look back at the derivative portal at Eq.~\eqref{DPDM}, it is actually a kinetic mixing term between the $Z$ boson and the $Z^{\prime}$ boson. For Abelian gauge bosons, one can write down the kinetic mixing term directly. For non-Abelian gauge bosons the kinetic mixing term can originate from loop corrections as shown in Fig.~\ref{fig:DPDMlo}, 
where $\Phi$ and $\Psi$ represent the scalar and fermion which contribute to the kinetic mixing respectively. The kinetic mixing can thus be estimated as~\cite{Holdom:1985ag,Rueter:2020qhf,Cai:2021evx} 
\begin{eqnarray}
    %\delta\epsilon\sim \frac{gg^{\prime}}{48\pi^2}(\ln (\frac{\mu^2}{m_{H}^2})+\ln (\frac{\mu^2}{m_{\Phi}^2}))=\frac{gg^{\prime}}{24\pi^2}\ln \frac{\mu^2}{m_{H}m_{\Phi}}
    \epsilon\sim \sum\limits_{i}  \frac{g_{i}g_{i}^{\prime}}{48\pi^2}\ln \frac{\mu^2}{m_{i}^2}-\sum\limits_{f}  \frac{g_{f}g_{f}^{\prime}}{12\pi^2}\ln \frac{\mu^2}{m_{f}^2},
    \label{kinmix}
\end{eqnarray}
where the first term and the second term represent contribution from scalars and fermions respectively, and $g_{i (f)},g_{i (f)}^{\prime}$ and $m_{i (f)}$ are the couplings and mass of the $i (f)$th particle which contribute to the kinetic mixing.  One thing should be kept in mind is that the kinetic mixing between photon and the massive gauge boson which couples to DM should be naturally small to not ruin the cancellation mechanism, which means the leading loop corrections to their kinetic mixing should be at least two-loop corrections as depicted by Fig.~\ref{fig:DPDMnlo}. Interpreting out the heavy particles that run in the loops of Fig.~\ref{fig:DPDMlo} and Fig.~\ref{fig:DPDMnlo} generates an effective action that contains not only the dimension-4 kinetic mixing operator $-\frac{\epsilon}{2}F^{\mu\nu}F_{\mu\nu}^{\prime}$ in Eq.~\eqref{DPDM}, but also higher-derivative operators such as $(\partial_{\rho}F^{\rho\mu}\partial^{\sigma}F^{\prime}_{\sigma\mu})/\Lambda^{2}$, which in momentum space scales as $k^{4}/\Lambda^{2}$~\cite{Gherghetta:2019coi}. This actually also falls back to the derivative portal, vanishing in the zero momentum transfer limit even when including $\frac{1}{k^{2}}$ from the photon's propagator. This means as long as the photon-$Z^{\prime}$ coupling term originated from two (or higher) loops diagram, the cancellation won't be spoiled by photon. In the explicit $\mathrm{SU}(2)_\mathrm{L}\times \mathrm{U}(1)_{Z^{\prime}}$ and $\mathrm{SU}(2)_\mathrm{L}\times \mathrm{SU}(2)_{Z^{\prime}}$ models discussed below, the fields $\Phi$ and $\psi_L$ that appear in Eq.~\eqref{Luv} and Eq.~\eqref{Luvt} are scalar and fermion which will mix with the Higgs boson and the neutrinos in SM, and they are exactly the particles that run in the loop of Fig.~\ref{fig:DPDMlo} and generate the derivative portal according to Eq.~\eqref{kinmix}.

\section{Building the Derivative Portal}\label{app:uv}%
Building the derivative portal is simple, while making a naturally small  kinetic mixing between the photon and the dark gauge boson which couples to DM is non-trivial (especially for the case where the SM $Z$ boson is in the derivative portal). Because generally one can write down a kinetic mixing term between the photon and another U(1) gauge boson directly. However, this can be avoided by assuming the kinetic mixing is in the same order of magnitude as its leading loop corrections or by embedding the dark gauge boson into a non-Abelian gauge group. In the following we will present three DPDM models and show the origination of their derivative portal and the kinetic mixing between the photon and the dark gauge boson. In the first two models,  we assume the kinetic mixing between two U(1) gauge bosons is in the same order of magnitude as its leading loop corrections, and we show that in these two models the kinetic mixing between the photon and the dark gauge boson is truly in two-loop corrections level. In the third model, we embed the dark gauge boson into a non-Abelian gauge group, and show that the kinetic mixing between the photon and the dark gauge boson originates from two-loop corrections level.

\subsection{The $\mathrm{U}(1)_\mathrm{B-L}\times \mathrm{U}(1)_\mathrm{X}$ model}
\label{sub:u1blu1x}

\begin{figure}[ht]
    \centering
    \includegraphics[width=0.4\textwidth]{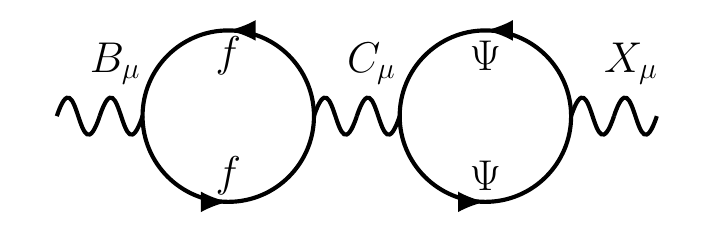}
    \caption{Kinetic mixing between $B_{\mu}$ and $X_{\mu}$.  }
    \label{fig:DPDMnlo}
\end{figure}

A simple and direct construction of the DPDM model is extending an $\mathrm{U}(1)_\mathrm{B-L}$ model with an extra $\mathrm{U}(1)_\mathrm{X}$ gauge symmetry. The relevant Lagrangian can be given by:
\begin{eqnarray}
    \label{model:2u1}
    \mathcal{L}= &&-\frac{1}{4}C^{\mu\nu}C_{\mu\nu}-\frac{1}{4}X^{\mu\nu}X_{\mu\nu}-\frac{\epsilon}{2}C^{\mu\nu}X_{\mu\nu}\\
	&&+\sum\limits_{f}  g_{BL}n_{f}C_{\mu}\bar{f}\gamma^{\mu}f+g_{X}X_{\mu}\bar{\chi}\gamma^{\mu}\chi\nonumber\\
	&&+\frac{1}{2}m_{C}^2C_{\mu}C^{\mu}+\frac{1}{2}m_{X}^2X_{\mu}X^{\mu}-m_{\chi}\bar{\chi}\chi\nonumber,
\end{eqnarray}
where $C$ and $X$ are gauge bosons of $\mathrm{U}(1)_\mathrm{B-L}$ and $\mathrm{U}(1)_\mathrm{X}$ symmetry\footnote{We also constructed a model which extends an $\mathrm{U}(1)_\mathrm{B-L}$ model with an $\mathrm{U}(1)_\mathrm{X}$ gauge symmetry in ~\cite{Cai:2021evx}. While in that construction the two extra gauge bosons are linked by mass mixing rather than kinetic mixing.}. The kinetic mixing term can be written directly or generated from loop corrections. To make it consistent we will consider all kinetic mixing comes from loop corrections. Therefore the derivative portal can be generated from the following Lagrangian:
\begin{eqnarray}
	\mathcal{L}= &&  g_{BL}C_{\mu}\bar{\Psi}\gamma^{\mu}\Psi+g_{X}X_{\mu}\bar{\Psi}\gamma^{\mu}\Psi.
\end{eqnarray}
With the above interactions the derivative portal can be generated through the second diagram in Fig.~\ref{fig:DPDMlo}. While kinetic mixing between the SM $B_{\mu}$ and $X_{\mu}$ is generated from two-loop corrections as displayed in Fig.~\ref{fig:DPDMnlo}. Because there is no particle coupling directly to both $B_{\mu}$ and $X_{\mu}$.
Note that there will also be kinetic mixing between $C_{\mu}$ and $B_{\mu}$, however this will not affect the cancellation mechanism since DM $\chi$ only couples to $X_{\mu}$. The constraints to the kinetic mixing between $C_{\mu}$ and $B_{\mu}$ can refer to that of the $\mathrm{U}(1)_\mathrm{B-L}$ theories~\cite{Emam:2007dy,Basso:2008iv,Basso:2010pe,Das:2021esm}.

\subsection{The $\mathrm{SU}(2)_\mathrm{L}\times \mathrm{U}(1)_{Z^{\prime}}$ model}
\label{sub:_z_boson_dpdm_model}

Another construction of the DPDM model is taking the SM $Z$ boson as one of the mediators in the derivative portal. The difficulty in this setting is that: since both the $Z$ boson and the photon are combinations of $B_{\mu}$ of $W^{3}_{\mu}$, it is not easy to make the extra gauge boson couples to the $Z$ boson at one-loop level while couples to the photon at two-loop level. Fortunately there are particles in the SM which couple to the $Z$ boson but the photon, which are the Higgs boson and neutrinos. Therefore we can use the neutrinos to generate a derivative portal between the SM $Z$ boson and an extra gauge boson $Z^{\prime}$, while leaving the photon out of the picture. The $Z^{\prime}$ is a gauge boson of an $\mathrm{U}(1)_{Z^{\prime}}$ gauge field, which can also be denoted as $\mathrm{U}(1)_{X}$.  One possible UV complete model of the DPDM model can thus be written as:
\begin{eqnarray}\label{Luv}
	    \mathcal{L}&&=\mathcal{L}_{SM}+(D^{\mu}\Phi)^{\dagger}D_{\mu}\Phi+\mu_{\Phi}^2|\Phi|^2-\lambda_{\Phi}|\Phi|^4+\lambda_{H\Phi}|H|^2|\Phi|^2\nonumber\\
		       &&-\frac{1}{4}Z^{\prime\mu\nu}Z^{\prime}_{\mu\nu}+i\bar{\chi}\gamma^{\mu}D_{\mu}\chi -m_{\chi}\bar{\chi}\chi+i\overbar{\psi_{L}}\gamma^{\mu}D_{\mu}\psi_{L}\\
		       &&+i\overbar{N_{R}}\slashed{\partial}N_{R}-\frac{1}{2}M_{N}\overbar{N_{R}^{c}}N_{R}-Y_{\nu}\tilde{H}\overbar{L_{L}}N_{R}-Y_{\psi}\Phi\overbar{\psi_{L}}N_{R}+\mathrm{h.c.}\nonumber,
\end{eqnarray}
where $\chi$ and $\Phi$ are DM and dark scalar respectively, $L$ is the SM lepton doublet (it can also be an extra fermion doublet), and $N_{R}$ is a right-handed ``neutrino" that will give mass to either the extra fermion $\psi_{L}$ or the $L$'s neutral component $\nu_{L}$\footnote{Note a new $\mathrm{U}(1)$ gauge symmetry and a left-handed fermion will induce anomaly. To cancel the anomaly we need to introduce an right-handed fermion which possesses the same $\mathrm{U}(1)$ charge as the left-handed fermion. The right-handed fermion can be neglected by chosing very large mass and a global $\mathrm{U}(1)$ symmetry.}. The covariant derivatives are given by:
\begin{eqnarray}
    &&D_{\mu}\Phi=(\partial_{\mu}-ig_{\chi}Z_{\mu}^{\prime})\Phi\nonumber\\ 
    &&D_{\mu}\chi=(\partial_{\mu}-ig_{\chi}n_{\chi}Z_{\mu}^{\prime})\chi\\ 
    &&D_{\mu}\psi_{L}=(\partial_{\mu}-ig_{\chi}Z_{\mu}^{\prime})\psi_{L}\nonumber,
\end{eqnarray}
where $g_{\chi}$ and $n_{\chi}$ are the gauge coupling and the quantum number of DM $\chi$. After $H$ and $\Phi$ get their vacuum expectation value $v_{H}$ and $v_{\Phi}$, we can write the mass matrix of $\nu_{L},\ N_{R}$ and $\psi_{L}$ as:    
\begin{eqnarray}
		       \frac{1}{2}\begin{pmatrix}
			   0&Y_{\nu}v_{H}&0\\
			    Y_{\nu}v_{H}&M_{N}&Y_{\psi}v_{\Phi}\\
			   0&Y_{\psi}v_{\Phi}&0
		       \end{pmatrix}\label{massnv}.
\end{eqnarray}
In these three particles the $Z$ boson couples to $\nu_{L}$ and the $Z^{\prime}$ boson couples to $\psi_{L}$, therefore after diagonalizing these three particles to their mass eigenstates, they all couple to the $Z$ and $Z^{\prime}$ bosons simultaneously, without coupling to the photon. Thus these particles can generate the kinetic mixing between the $Z$ and $Z^{\prime}$ bosons through one-loop corrections. While the kinetic mixing between the photon and the $Z^{\prime}$ boson are generated through two-loop corrections, as illustrated in Fig.~\ref{fig:DPDMnlo} (with $X_{\mu},\ C_{\mu},$ and $B_{\mu}$ replaced by $Z^{\prime}_{\mu},\ Z_{\mu},$ and $A_{\mu}$, and $\Psi$ replaced by neutrinos). Note that with a large $v_{\Phi}$ the $Z^{\prime}\bar{f}f$ coupling originated from mixing between $H$ and $\Phi$ can be neglected, while for completion we add the kinetic mixing origination from scalars in Appendix~\ref{app:scalarKM}. In the loop computation of the mixing between $Z$ and $Z^{\prime}$, the mixed two-point function is transverse and can be written as $\Pi_{ZZ^{\prime}}^{\mu\nu}(q)=(g^{\mu\nu}q^{2}-q^{\mu}q^{\nu})\Pi_{ZZ^{\prime}}(q^{2})$. Expanding $\Pi_{Z Z^{\prime}}(q^{2})$ around $q^{2}=0$ generates momentum dependent kinetic mixing and higher derivative operators in the quadratic kernel of the vector bosons. And it will not introduce an independent non-derivative (dimension-2) mass mixing term between them. In our phenomenological analysis we work with the effective Lagrangian after including this loop generated kinetic mixing in the parameters that enter the quadratic terms, and we consistently apply the cancellation condition proven in Appendix~\ref{app:uproof} to the resulting mass eigenstates. In this setup all gauge-boson mediated contributions to the $\chi-$SM amplitude remain proportional to the momentum transfer in the direct detection regime.      

\subsection{The $\mathrm{SU}(2)_\mathrm{L}\times \mathrm{SU}(2)_{Z^{\prime}}$ model}
\label{sub:_su_2__l_}

To avoid the assumption that the kinetic mixing between the photon and the DM is in the same order of magnitude as its leading loop corrections, alternatively, we can embed the $Z^{\prime}$ boson in the $\mathrm{SU}(2)_\mathrm{L}\times \mathrm{U}(1)_{Z^{\prime}}$ model to be a member of a multiplet, then there will be no tree-level kinetic mixing between the photon and the $Z^{\prime}$ boson. For example, to embed $Z^{\prime}$ in a non-Abelian gauge group $\mathrm{SU}(2)_{Z^{\prime}}$, Eq.~\eqref{Luv} can be modified to:
\begin{eqnarray}\label{Luvt}
	    \mathcal{L}&&=\mathcal{L}_{SM}+(D^{\mu}\Phi)^{\dagger}D_{\mu}\Phi+\mu_{\Phi}^2|\Phi|^2-\lambda_{\Phi}|\Phi|^4+\lambda_{H\Phi}|H|^2|\Phi|^2\nonumber\\
		       &&-\frac{1}{4}W^{\prime\mu\nu}_{a}W^{\prime a}_{\mu\nu}+i\bar{\chi}\gamma^{\mu}D_{\mu}\chi -m_{\chi}\bar{\chi}\chi+i\overbar{\psi_{L}}\gamma^{\mu}D_{\mu}\psi_{L}\\
		       &&+i\overbar{N_{R}}\slashed{\partial}N_{R}-\frac{1}{2}M_{N}\overbar{N_{R}^{c}}N_{R}-Y_{\nu}\tilde{H}\overbar{L_{L}}N_{R}-Y_{\psi}\tilde{\Phi}\overbar{\psi_{L}}N_{R}+\mathrm{h.c.},\nonumber
\end{eqnarray}
where $\Phi,\ \chi$ and $\psi_{L}$ are $\mathrm{SU}(2)_{Z^{\prime}}$ doublet now (with $\tilde{\Phi}=i\sigma^{2}\Phi^{*}$ being an analog to $\tilde{H}=i\sigma^{2}H^{*}$), and the $Z^{\prime}$ boson becomes a neutral component of the $\mathrm{SU}(2)_{Z^{\prime}}$ gauge filed $W^{\prime a}_{\mu}$. Therefore the covariant derivatives are given by: 
\begin{eqnarray}
    &&D_{\mu}\Phi=(\partial_{\mu}-ig_{\chi}W_{\mu}^{\prime a}\tau^{a})\Phi\nonumber\\ 
    &&D_{\mu}\chi=(\partial_{\mu}-ig_{\chi}n_{\chi}W_{\mu}^{\prime a}\tau^{a})\chi\\ 
    &&D_{\mu}\psi_{L}=(\partial_{\mu}-ig_{\chi}W_{\mu}^{\prime a}\tau^{a})\psi_{L}\nonumber.
\end{eqnarray}
Compared to Eq.~\eqref{Luv}, there are more particles in Eq.~\eqref{Luvt}. For example, there are two DMs in Eq.~\eqref{Luvt} (i.e., both components of $\mathrm{SU}(2)_{Z^{\prime}}$ doublet $\chi$ are DM). These two DMs couple to the $Z^{\prime}$ boson in the same way, and the direct detection cancellation works for both DMs since the building of the derivative portal is the same as that in the $\mathrm{SU}(2)_\mathrm{L}\times \mathrm{U}(1)_{Z^{\prime}}$ model. In this model there is no tree-level kinetic mixing between the photon and the $Z^{\prime}$ boson. Their kinetic mixing originates from two-loop corrections, as illustrated in Fig.~\ref{fig:DPDMnlo} (with $X_{\mu},\ C_{\mu},$ and $B_{\mu}$ replaced by $Z^{\prime}_{\mu},\ Z_{\mu},$ and $A_{\mu}$, and $\Psi$ replaced by neutrinos). Just like that in the SM, the imaginary part of the neutral component of $\Phi$ and the charged component of $\Phi$ will be eaten by the $Z^{\prime}$ boson and the $W^{\prime \pm}$ bosons respectively, and thus giving mass to these gauge bosons. Note that the mass of the $W^{\prime \pm}$ bosons will be exactly the same as the mass of the $Z^{\prime}$ boson, which is safe since the charge the $W^{\prime \pm}$ bosons possess is not the same as the electric charge in the SM. To keep the $W^{\prime \pm}$ bosons not involved in the derivative portal, we have imposed a global U(1) symmetry to Eq.~\eqref{Luvt}, under which $\Phi$ and $\psi_{L}$ charged oppositely. This symmetry will prohibit the $\Phi \overbar{\psi_{L}}N_{R}$ term, and thus prohibiting the charged component of $\psi_{L}$ get into mass matrix like Eq.~\eqref{massnv}. Therefore the $W^{\prime \pm}$ bosons will not couple to the fermions which couple to the SM $Z$ boson. Also one might note that the Higgs portal can also generate DM-SM fermions interactions, however the Higgs portal can always be neglected naturally by setting the dark scalar mass much heavier than the Higgs boson mass. Large dark scalar mass will lead to negligible mixing between the dark scalar and the Higgs boson, thus keeping the Higgs portal influence out of the picture.

\section{DM Phenomenology}\label{sec:phe}

In this section we will study the DM phenomenology, which includes exploring constraints from observed DM relic density, DM indirect detection, and collider search. We focus on a effective DPDM model where the SM $Z$ boson and a dark boson $Z^{\prime}$ are chosen as mediators, then the relevant Lagrangian can be written as:
\begin{eqnarray}
    \mathcal{L}=&&-\frac{1}{4}Z^{\mu\nu}Z_{\mu\nu}-\frac{1}{4}Z^{\prime\mu\nu}Z^{\prime}_{\mu\nu}-\frac{\epsilon}{2} Z^{\mu\nu}Z_{\mu\nu}^{\prime}\\
	&&+\sum\limits_{f}  Z_{\mu}\bar{f}\gamma^{\mu}(g_{V}-g_{A}\gamma^{5})f+g_{\chi}Z_{\mu}^{\prime}\bar{\chi}\gamma^{\mu}\chi\nonumber\\
	&&+\frac{1}{2}m_{Z}^2Z_{\mu}Z^{\mu}+\frac{1}{2}m_{Z^{\prime}}^2Z_{\mu}^{\prime}Z^{\prime\mu}-m_{\chi}\bar{\chi}\chi\nonumber,
\end{eqnarray}
where the first, the second, and the third lines represent the kinetic terms, the coupling terms, and the mass terms respectively. Since the kinetic mixing term comes from loop corrections, this Lagrangian is not UV complete. The possible UV complete models can be seen in Sec.~\ref{app:uv}.

To normalize the kinetic terms, we can apply the following transformation:
\begin{eqnarray}
	\begin{pmatrix}
	    Z_{\mu}\\
	    Z_{\mu}^{\prime}
	\end{pmatrix}=\frac{1}{\sqrt{2} }\begin{pmatrix}
	-\frac{1}{\sqrt{1-\epsilon} }&\frac{1}{\sqrt{1+\epsilon} }\\
	\frac{1}{\sqrt{1-\epsilon} } &\frac{1}{\sqrt{1+\epsilon} }
	\end{pmatrix}\begin{pmatrix}
	    \tilde{Z}_{\mu}\\
	    \tilde{Z}_{\mu}^{\prime} 
	\end{pmatrix}.
\end{eqnarray}
After this operation there will be a mixing term in the mass matrix of $\tilde{Z} $ boson and $\tilde{Z}^{\prime} $ boson. After diagonalizing these bosons to their mass eigenstates, one can prove the cancellation in the usual way. See Appendix~\ref{app:uproof} for more details. 

\begin{figure*}[ht]
    \centering
    \includegraphics[width=0.45\textwidth]{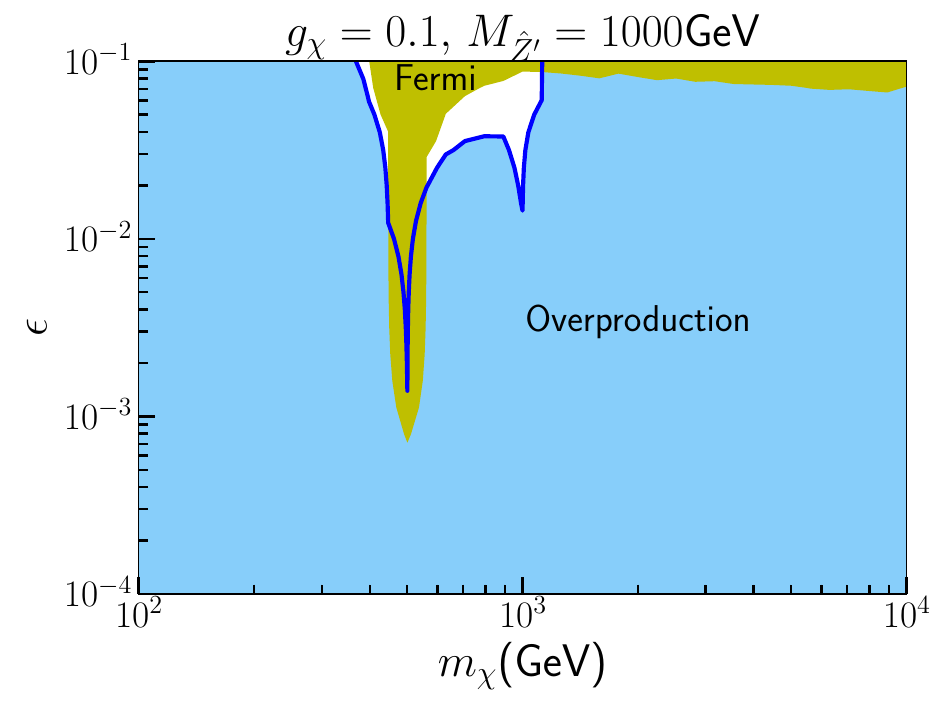}
    \includegraphics[width=0.45\textwidth]{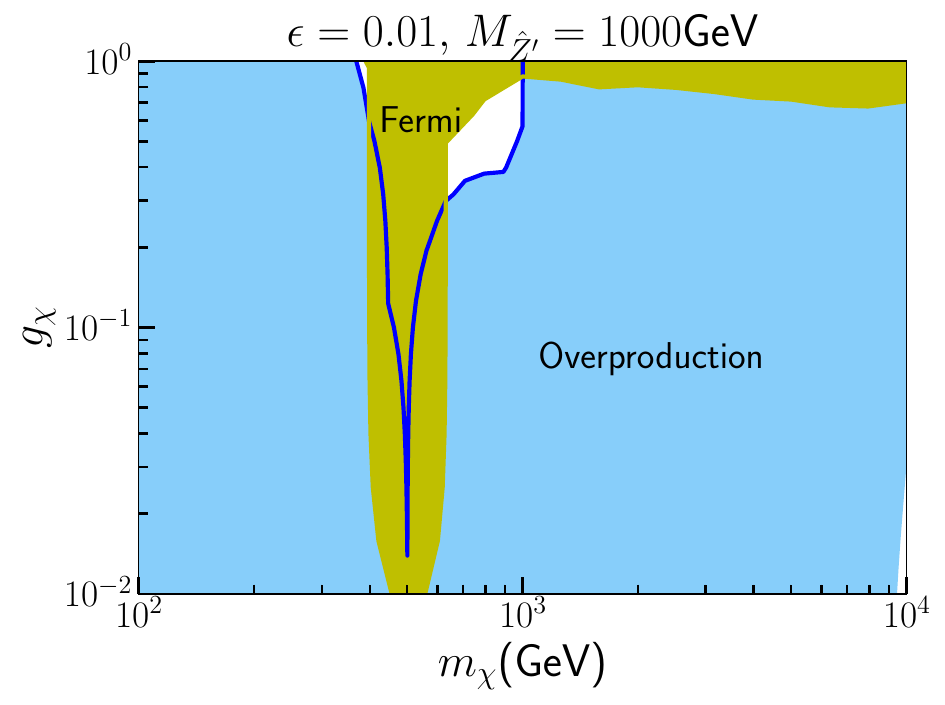}
    \includegraphics[width=0.45\textwidth]{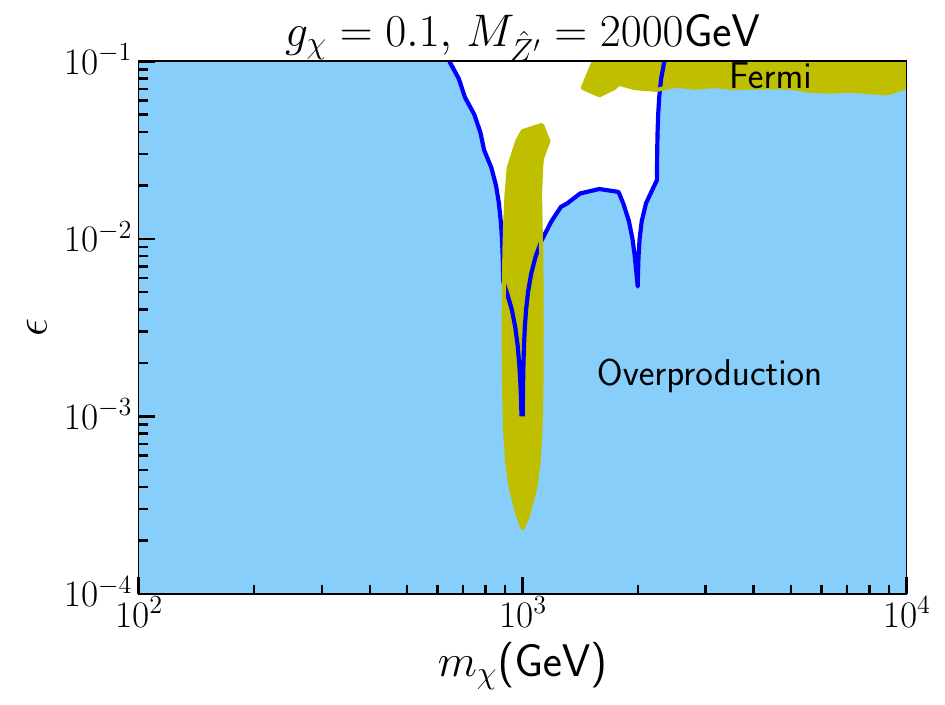}
    \includegraphics[width=0.45\textwidth]{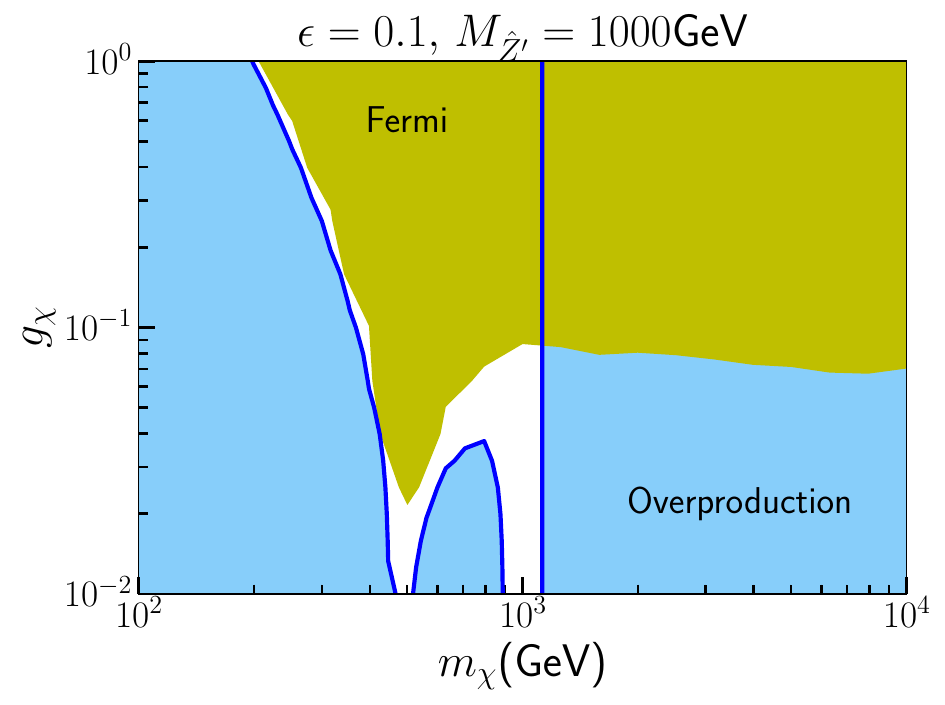}
    \caption{Constraints on the $\mathrm{SU}(2)_\mathrm{L}\times \mathrm{U}(1)_{Z^{\prime}}$ model. The blue lines are contours that saturate the Planck experiment~\cite{Planck:2018vyg} observation of DM relic density, while the lightblue area are excluded by the Planck experiment, and the yellow area are excluded by the DM indirect detection bounds from Fermi-LAT experiment~\cite{Ambrogi:2018jqj,Fermi-LAT:2016uux} .}
    \label{fig:res}
\end{figure*}

We implemented this model in \texttt{FeynRules~2}~\cite{Alloul:2013bka}, and utilized the \texttt{MadGraph}~\cite{Alwall:2014hca} plugin \texttt{MadDM}~\cite{Ambrogi:2018jqj} to calculate the DM relic density and the DM indirect detection bounds. The results are shown in Fig.~\ref{fig:res},
where the blue lines are contours that saturate the Planck experiment~\cite{Planck:2018vyg} observation of DM relic density, while the lightblue area are excluded by the Planck experiment, and the yellow area are excluded by the DM indirect detection bounds from Fermi-LAT experiment~\cite{Ambrogi:2018jqj,Fermi-LAT:2016uux} . There are four free parameters in the DPDM model, which are $m_{\chi},\ \epsilon,\ g_{\chi}$ and $m_{\hat{Z}^{\prime}}$ (the mass of mass eigenstates of $Z^{\prime}$), with the measured SM $Z$ boson mass being an input parameter. We adopt the DM mass $m_{\chi}$ vs the kinetic mixing coupling $\epsilon$ and the gauge coupling $g_{\chi}$ in the left and right panel of Fig.~\ref{fig:res} respectively. In the left panel of Fig.~\ref{fig:res} we fix $g_{\chi}=0.1$, and set the $Z^{\prime}$ mass in the upper and lower figure to be $m_{\hat{Z}^{\prime}}=1000~\mathrm{GeV}$ and $m_{\hat{Z}^{\prime}}=2000~\mathrm{GeV}$ respectively. In the right panel of Fig.~\ref{fig:res} we fix $m_{\hat{Z}^{\prime}}=1000~\mathrm{GeV}$, and set $\epsilon$ in the upper and lower figure to be $\epsilon=0.01$ and $\epsilon=0.1$ respectively. In Fig.~\ref{fig:res} the blue lines plus the white area are viable parameters space of DM. There are dips which correspond to the resonant annihilation that happens when DM is around half of $m_{\hat{Z}^{\prime}}$. For DM relic density there are also dips around $m_{\hat{Z}^{\prime}}$ which are caused by DM coannihilating with the dark mediator $Z^{\prime}$. For DM indirect detection there is no coannihilation since nowadays there is no $Z^{\prime}$ in space. 

From Fig.~\ref{fig:res} we see that the kinetic mixing coupling should be in the order similar or larger than $O(0.01)$ to not being constrained severely by DM relic density. Larger $\epsilon$ or $g_{\chi}$ will result in larger relic density since larger kinetic mixing or coupling represents larger interaction and hence leads to larger annihilation cross section of DM. When the DM mass is around half of the mass of the dark vector boson or is about the same as the mass of the dark vector boson, there will be resonant annihilation or coannihilation which will strongly enhance the annihilation cross section of DM. Thanks to the coannihilation there are viable parameters space which are not excluded by DM indirect detection. Since the couplings and mass of both DM and the dark vector boson are free parameters, this leaves large parameters space for future phenomenology studies~\cite{PerezdelosHeros:2020qyt,Zeng:2021moz}.

Since the $Z^{\prime}$ couples to SM fermions through mixing with the $Z$ boson. It might be constraint by the collider search for new gauge boson coupling to leptons. This constraints from collider like LEP~\cite{ALEPH:2006bhb,ALEPH:2013dgf} can be formulated as~\cite{Arcadi:2024obp,Falkowski:2015krw}:  
\begin{eqnarray}
    g_{V}<\frac{0.12M_{Z^{\prime}}}{\mathrm{TeV}},\ g_{A}<\frac{0.19M_{Z^{\prime}}}{\mathrm{TeV}},
    \label{gvac}
\end{eqnarray}
where $M_{Z^{\prime}}$ represents the mass of $Z^{\prime}$ and $g_{V}$ and $g_{A}$ represent the vector and axial coupling between $Z^{\prime}$ and SM leptons respectively. Compared to LEP, constraints from LHC are much stronger for large mass of $Z^{\prime}$~\cite{Arcadi:2024obp,ATLAS:2019erb}, and we extracted the LHC bounds in Table~\ref{tab:gvga}. For our model we have calculated the corresponding upper limit of the coupling and some typical couplings as listed in Table~\ref{tab:gvga}. In Table~\ref{tab:gvga} the above three rows are the values calculated directly from our model and the last two rows are the LHC bounds and the upper limit of $\epsilon$ calculated from the $g_A$ constraints. The upper limits of $\epsilon$ for $M_{\hat{Z}^{\prime}}=1~\mathrm{TeV}$ and $M_{\hat{Z}^{\prime}}=5~\mathrm{TeV}$ are $0.868$ and $0.983$ respectively.    
\begin{table}[ht]
    \centering
    \begin{tabular}{cccc}
	\hline
	mass&coupling&$g_V$&$g_A$ \\
	$M_{\hat{Z}^{\prime}}=1~\mathrm{TeV}$&$ \epsilon=10^{-3}$&$1.03\times 10^{-7}$&$1.55\times 10^{-6}$ \\ 
    $M_{\hat{Z}^{\prime}}=1~\mathrm{TeV}$&$ \epsilon=10^{-2}$&$1.03\times 10^{-6}$&$1.55\times 10^{-5}$ \\ 
    $M_{\hat{Z}^{\prime}}=1~\mathrm{TeV}$&$ \epsilon=10^{-1}$&$1.04\times 10^{-5}$&$1.57\times 10^{-4}$\\
	\hline
	mass&upper $\epsilon$ &upper $g_V$&upper $g_A$ \\
	$M_{\hat{Z}^{\prime}}=1~\mathrm{TeV}$&$ 0.868$&$6.54\times 10^{-3}$&$5.69\times 10^{-3}$ \\ 
	$M_{\hat{Z}^{\prime}}=5~\mathrm{TeV}$&$ 0.983$&$0.172$&$0.150$ \\ 
	\hline
    \end{tabular}
    \caption{Information of couplings between $Z^{\prime}$ and SM leptons.}
    \label{tab:gvga}
\end{table}
From Eq.~\eqref{gvac} and Table~\ref{tab:gvga} we see that our model is safe from the constraints of couplings between $Z^{\prime}$ boson and the SM leptons.   

\begin{figure}[ht]
    \centering
    \includegraphics[width=0.4\textwidth]{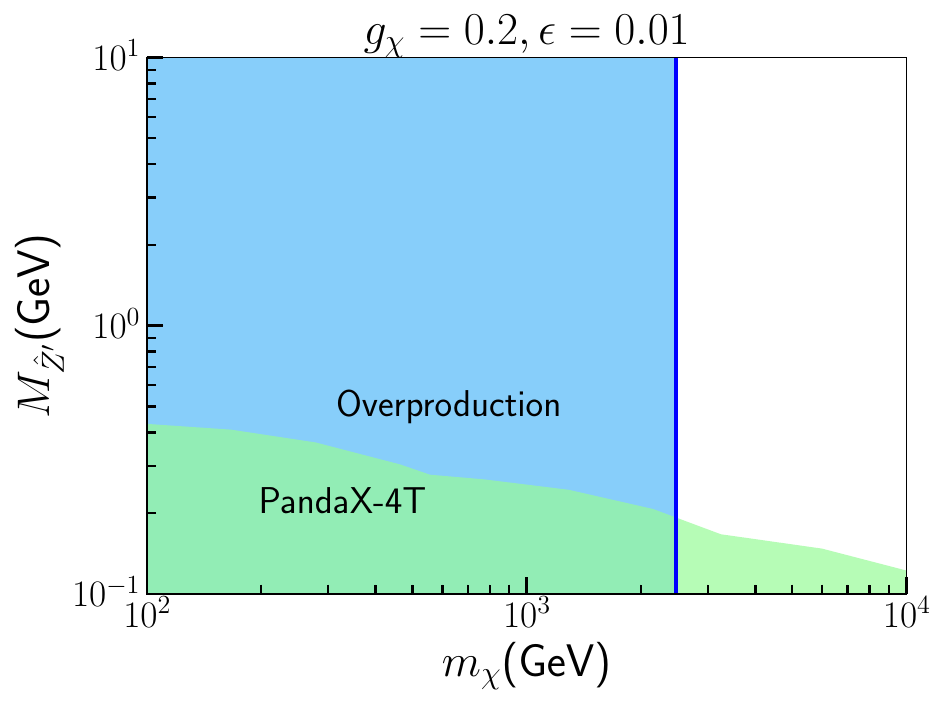}
    \caption{Direct detection and relic density constrains on the DPDM model with small $M_{\hat{Z}^{\prime}}$.}
    \label{fig:dd}
\end{figure}

To show the current DM direct detection on the DPDM model, we explored small $M_{\hat{Z}^{\prime}}$ area since the zero momentum transfer limit will break when the mediator mass $M_{\hat{Z}^{\prime}}$ approaches the transferred momentum. To handle the non-zero momentum transfer limit case in DM direct detection we have to calculate and compare the scattering events rather than comparing DM-necleuon scattering cross section~\cite{Zeng:2021moz}. The result are show in Fig.~\ref{fig:dd}, where the green area are excluded by DM direct detection experiment PandaX-4T~\cite{PandaX-4T:2021bab} and the lightblue area are excluded by the Planck experiment~\cite{Planck:2018vyg}. 

\begin{figure}[ht]
    \centering
    \includegraphics[width=0.4\textwidth]{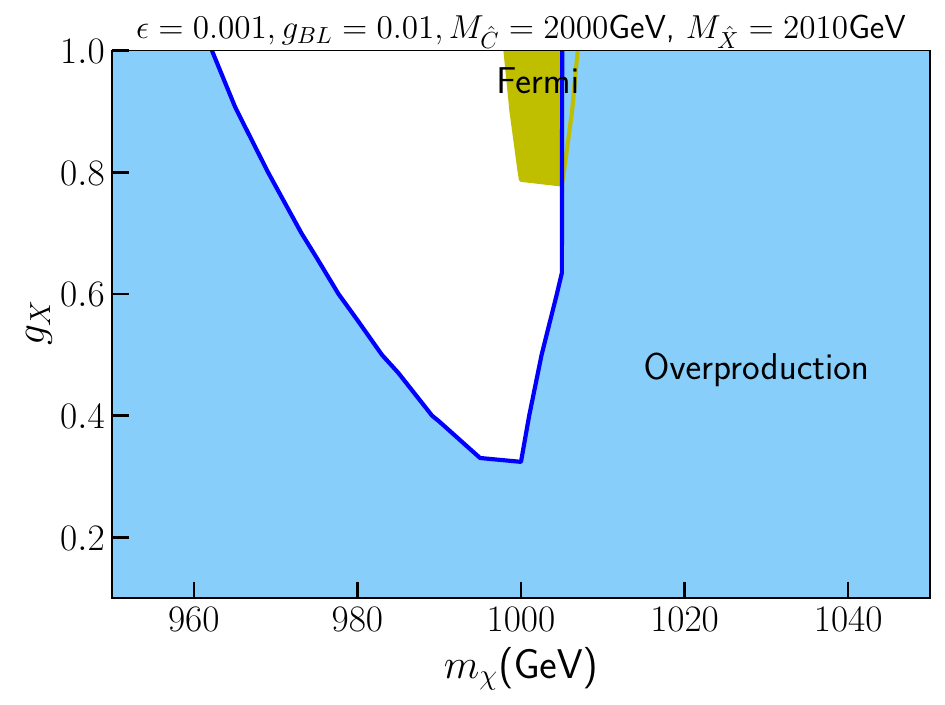}
    \includegraphics[width=0.4\textwidth]{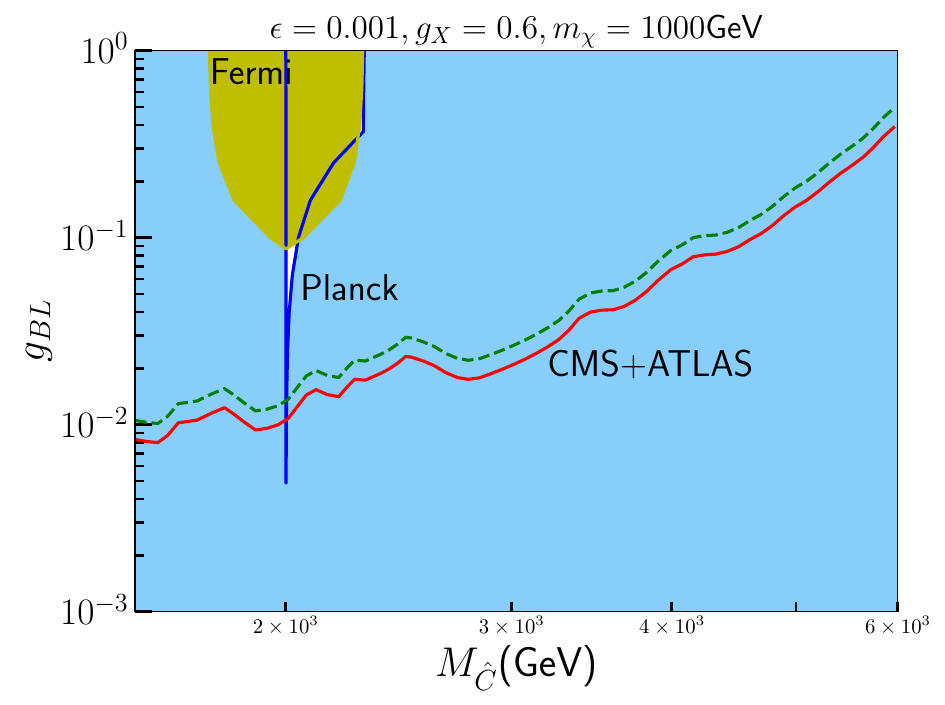}
    \caption{Constraints on the $\mathrm{U}(1)_\mathrm{B-L}\times \mathrm{U}(1)_\mathrm{X}$ model.}
    \label{fig:u1bl}
\end{figure}

For demonstration purpose we add Eq.~\eqref{model:2u1} upon a $\mathrm{U}(1)_\mathrm{B-L}$ model~\cite{Das:2021esm} and implemented the complete $\mathrm{U}(1)_\mathrm{B-L}\times \mathrm{U}(1)_\mathrm{X}$ model through the same procedure as that of the above model. The results are show in Fig.~\ref{fig:u1bl}, where $M_{\hat{C}}, M_{\hat{X}}$ are the diagonalized masses of $C_{\mu}$ and $X_{\mu}$, and $n_f$ in Eq.~\eqref{model:2u1} represents the $B-L$ number of the corresponding fermion. 
In Fig.~\ref{fig:u1bl} the lightblue area is excluded by observed DM relic density, and the blue lines correspond to the observed DM relic density. The yellow area is excluded by DM indirect detection. The area above the red line is excluded by LHC (CMS~\cite{CMS:2019tbu} plus ATLAS~\cite{ATLAS:2019erb}) constraints on the $Z^{\prime}ll$ couplings~\cite{Das:2021esm}, with degenerate masses of $C_{\mu}$ and $X_{\mu}$ assumed. As a comparison, the dashed green line shows the LHC constraints on the $\mathrm{U}(1)_\mathrm{B-L}$ model~\cite{Das:2021esm}. In the upper figure of Fig.~\ref{fig:u1bl}, $g_{BL}=0.01$ is chosen to be slightly smaller than it's upper limit $0.0106$ given by the LHC constraint, and the white space are the viable parameters space. The lower figure of Fig.~\ref{fig:u1bl} expanded the surived point ($m_{\chi}=1000~\mathrm{GeV},g_X=0.6,\epsilon=0.001, g_{BL}=0.01, M_{\hat{C}}=2000~\mathrm{GeV}, M_{\hat{X}}=2010~\mathrm{GeV}$ ) from the upper figure in $M_{\hat{C}}$(with $M_{\hat{X}}=M_{\hat{C}}+10~\mathrm{GeV}$) and $g_{BL}$, and we see the blue line below the red line are the viable parameters space.  

\section{Conclusion and Discussion} \label{sec:con}
In this work we have proposed a new kind of DM model: the DPDM model. We proved this kind of model possesses cancellation mechanism and thus escapes stringent DM direct detection constraints naturally. We explored the UV origination of this kind of model and constructed three UV complete DPDM models where the derivative port originate naturally. Furthermore, we have also studied the DM relic density predicted by the DPDM models as well as constraints from DM direct detection, DM indirect detection and collider searches. We found the models survive from all these constraints.

In this work we focused on the framework of the DPDM model, while the derivative portal can link to a vast variety of DM models. For example, one can easily implement UV complete DPDM models with two extra massive U(1) gauge bosons by coupling these two vector bosons to a same heavy particle. Though it is not easy to do the same for the SM $Z$ boson and an extra vector boson, in Sec.~\ref{app:uv} we adopt neutrino to build the derivative portal and presented two possible UV complete models. In these constructions, the dark sector is not deeply involved, and a mirror world might be preferred in the dark sector of these UV complete DPDM models~\cite{Koren:2019iuv,Alizzi:2021vyc}. 

Also there are works can be studied in the future: the possible UV completion model in Sec.~\ref{app:uv} might be able to give mass to neutrinos and phenomenology studies like electroweak oblique parameters constraints. 

%\begin{acknowledgments}
%    acknowledgments
%\end{acknowledgments}
\appendix

\section{Proof of cancellation mechanism in mass eigenstates}\label{app:uproof}
In the proof of cancellation mechanism in the DPDM model, the mass term of the DM is irrelevant. Therefore the relevant Lagrangian reads:
\begin{eqnarray}
    \mathcal{L}=&&-\frac{1}{4}Z^{\mu\nu}Z_{\mu\nu}-\frac{1}{4}Z^{\prime\mu\nu}Z^{\prime}_{\mu\nu}-\frac{\epsilon}{2} Z^{\mu\nu}Z_{\mu\nu}^{\prime}\\
	&&+\sum\limits_{f}  Z_{\mu}\bar{f}\gamma^{\mu}(g_{V}-g_{A}\gamma^{5})f+g_{\chi}Z_{\mu}^{\prime}\bar{\chi}\gamma^{\mu}\chi\nonumber\\
	&&+\frac{1}{2}m_{Z}^2Z_{\mu}Z^{\mu}+\frac{1}{2}m_{Z^{\prime}}^2Z_{\mu}^{\prime}Z^{\prime\mu}\nonumber.
\end{eqnarray}
After the following transformation
\begin{eqnarray}
	\begin{pmatrix}
	    Z_{\mu}\\
	    Z_{\mu}^{\prime}
	\end{pmatrix}=\frac{1}{\sqrt{2} }\begin{pmatrix}
	-\frac{1}{\sqrt{1-\epsilon} }&\frac{1}{\sqrt{1+\epsilon} }\\
	\frac{1}{\sqrt{1-\epsilon} } &\frac{1}{\sqrt{1+\epsilon} }
	\end{pmatrix}\begin{pmatrix}
	    \tilde{Z}_{\mu}\\
	    \tilde{Z}_{\mu}^{\prime} 
	\end{pmatrix},
\end{eqnarray}
the kinetic terms are normalized and the Lagrangian becomes
%\begin{widetext}
%\begin{eqnarray}
%    \mathcal{L}=&&-\frac{1}{4}\tilde{Z}^{\mu\nu}\tilde{Z}_{\mu\nu}-\frac{1}{4}\tilde{Z}^{\prime\mu\nu}\tilde{Z}^{\prime}_{\mu\nu}+\sum\limits_{f}  \frac{1}{\sqrt{2} }(-\frac{1}{\sqrt{1-\epsilon} }\tilde{Z}_{\mu}+ \frac{1}{\sqrt{1+\epsilon} }\tilde{Z}^{\prime}_{\mu} )\bar{f}\gamma^{\mu}(g_{V}-g_{A}\gamma^{5})f+g_{\chi}\frac{1}{\sqrt{2} }(\frac{1}{\sqrt{1-\epsilon} }\tilde{Z}_{\mu}+ \frac{1}{\sqrt{1+\epsilon} }\tilde{Z}^{\prime}_{\mu} )\bar{\chi}\gamma^{\mu}\chi\nonumber\\
%	&&+\frac{1}{2}m_{Z}^2(\frac{1}{\sqrt{2} }(-\frac{1}{\sqrt{1-\epsilon} }\tilde{Z}_{\mu}+ \frac{1}{\sqrt{1+\epsilon} }\tilde{Z}^{\prime}_{\mu} ))^2+\frac{1}{2}m_{Z^{\prime}}^2(\frac{1}{\sqrt{2} }(\frac{1}{\sqrt{1-\epsilon} }\tilde{Z}_{\mu}+ \frac{1}{\sqrt{1+\epsilon} }\tilde{Z}^{\prime}_{\mu} ))^2.
%\end{eqnarray}
%\end{widetext}
%Define $k_1=1/\sqrt{2-2\epsilon} $ and $k_2=1/\sqrt{2+2\epsilon} $, then the Lagrangian will be simplified to:
    
\begin{eqnarray}
    \mathcal{L}=&&-\frac{1}{4}\tilde{Z}^{\mu\nu}\tilde{Z}_{\mu\nu}-\frac{1}{4}\tilde{Z}^{\prime\mu\nu}\tilde{Z}^{\prime}_{\mu\nu}+g_{\chi}(k_1\tilde{Z}_{\mu}+ k_2\tilde{Z}^{\prime}_{\mu} )\bar{\chi}\gamma^{\mu}\chi\nonumber\\
		&&+\sum\limits_{f}  (-k_1\tilde{Z}_{\mu}+ k_2\tilde{Z}^{\prime}_{\mu} )\bar{f}\gamma^{\mu}(g_{V}-g_{A}\gamma^{5})f\\
		&&+\frac{1}{2}m_{Z}^2(-k_1\tilde{Z}_{\mu}+ k_2\tilde{Z}^{\prime}_{\mu} )^2+\frac{1}{2}m_{Z^{\prime}}^2(k_1\tilde{Z}_{\mu}+ k_2\tilde{Z}^{\prime}_{\mu} )^2\nonumber,
\end{eqnarray}
where $k_1=1/\sqrt{2-2\epsilon} $ and $k_2=1/\sqrt{2+2\epsilon} $.
Then the mass matrix of the vector mediators can be written as:
\begin{eqnarray}
	&&\frac{1}{2}\begin{pmatrix}
	    \tilde{Z}_{\mu}&\tilde{Z}^{\prime}_{\mu}\end{pmatrix}OO^T\begin{pmatrix}
	    k_1^2M_1&k_1k_2M_2\\
	    k_1k_2M_2&k_2^2M_1
	    \end{pmatrix}OO^T\begin{pmatrix}
	        \tilde{Z}_{\mu}\\
		\tilde{Z}^{\prime}_{\mu}
	    \end{pmatrix}\nonumber\\
			   &&=
	\frac{1}{2}\begin{pmatrix}
	    \hat{Z}_{\mu}&\hat{Z}^{\prime}_{\mu}\end{pmatrix}\begin{pmatrix}
	    m_{\hat{Z} }^2&0\\
	    0&m_{\hat{Z}^{\prime} }^2
	    \end{pmatrix}\begin{pmatrix}
	        \hat{Z}_{\mu}\\
		\hat{Z}^{\prime}_{\mu}
	\end{pmatrix}\label{massmatrix},
\end{eqnarray}
where we have defined $M_1=m_{Z}^2+m_{Z^{\prime}}^2,\ M_2=m_{Z^{\prime}}^2-m_{Z}^2$, and $O$ is an orthogonal matrix that diagonalizes the mass matrix. $O$ can be defined as:
\begin{eqnarray}
	    O=\begin{pmatrix}
		\cos \theta&\sin \theta\\
		-\sin \theta& \cos \theta
	    \end{pmatrix}, 
\end{eqnarray}
with $\tan 2\theta$ being formulated as:
\begin{eqnarray}
    \tan 2\theta=\frac{2k_1k_2M_2}{(k_2^2-k_1^2)M_1}.
\end{eqnarray}
After diagonalization, the Lagrangian becomes
    
\begin{eqnarray}
    \mathcal{L}=&&-\frac{1}{4}\hat{Z}^{\mu\nu}\hat{Z}_{\mu\nu}-\frac{1}{4}\hat{Z}^{\prime\mu\nu}\hat{Z}^{\prime}_{\mu\nu}
    +\frac{1}{2}m_{\hat{Z}}^2\hat{Z}_{\mu}^2+\frac{1}{2}m_{\hat{Z}^{\prime}}^2\hat{Z}^{\prime 2}_{\mu}\nonumber\\
	&&+\sum\limits_{f}  \left((-k_2\sin \theta -k_1\cos \theta) \hat{Z}_{\mu}+(-k_1\sin \theta    +k_2\cos \theta) \hat{Z}^{\prime}_{\mu}   \right)\bar{f}\gamma^{\mu}(g_{V}-g_{A}\gamma^{5})f\\
	&&+g_{\chi}\left((k_1\cos \theta - k_2\sin \theta) \hat{Z}_{\mu} +(k_2\cos \theta+k_1\sin \theta) \hat{Z}^{\prime}_{\mu}   \right)\bar{\chi}\gamma^{\mu}\chi\nonumber.
\end{eqnarray}
With this Lagrangian we can write the scattering amplitude between the SM fermions and DM as:
    
\begin{eqnarray}
    i\mathcal{M}=&&(-i)^2\bar{u}(p_3)(-\gamma^{\mu}(g_{V}-g_{A}\gamma^{5}))u(p_1)\bar{u}(p_4)(-g_{\chi}\gamma^{\nu})u(p_2)\nonumber\\
		 &&\times \left(  \frac{-ig_{\mu\nu}(-k_2\sin \theta -k_1\cos \theta)(k_1\cos \theta - k_2\sin \theta)}{t-m_{\hat{Z}}^2}+\frac{-ig_{\mu\nu}(-k_1\sin \theta    +k_2\cos \theta)(k_2\cos \theta+k_1\sin \theta)}{t-m_{\hat{Z}^{\prime}}^2}\right)\nonumber\\
		 \propto&&\left(  \frac{(-k_2\sin \theta -k_1\cos \theta)(k_1\cos \theta - k_2\sin \theta)}{t-m_{\hat{Z}}^2}+\frac{(-k_1\sin \theta    +k_2\cos \theta)(k_2\cos \theta+k_1\sin \theta)}{t-m_{\hat{Z}^{\prime}}^2}\right)\nonumber\\
    =&&\frac{t(\ldots) }{(t-m_{\hat{Z}}^2)(t-m_{\hat{Z}^{\prime}}^2)}-\frac{m_{\hat{Z}^{\prime}}^2(k_2^2\sin ^2\theta-k_1^2\cos ^2\theta)+m_{\hat{Z}}^2(k_2^2\cos ^2\theta-k_1^2\sin ^2\theta) }{(t-m_{\hat{Z}}^2)(t-m_{\hat{Z}^{\prime}}^2)}\nonumber.
\end{eqnarray}
In the result of the above equation we have extracted the key structure of the cancellation mechanism. If the amplitude is proportional to the momentum transfer $t$, then it goes to zero in the zero momentum transfer limit. Thus the cancellation is valid when the last line of the above equation equals to zero. Which means:
\begin{eqnarray}
    \frac{m_{\hat{Z}^{\prime}}^2(k_2^2\sin ^2\theta-k_1^2\cos ^2\theta)+m_{\hat{Z}}^2(k_2^2\cos ^2\theta-k_1^2\sin ^2\theta) }{(t-m_{\hat{Z}}^2)(t-m_{\hat{Z}^{\prime}}^2)}=0.\nonumber
\end{eqnarray}
This equation is equivalent to
\begin{eqnarray}
\frac{k_2^2\sin ^2\theta-k_1^2\cos ^2\theta}{k_2^2\cos ^2\theta-k_1^2\sin ^2\theta}=-\frac{m_{\hat{Z} }^2}{m_{\hat{Z}^{\prime} }^2},
\end{eqnarray}
and we will prove that this equation is true. From Eq.~\eqref{massmatrix} we can write:
\begin{eqnarray}
    \frac{- m_{\hat{Z} }^2}{m_{\hat{Z}^{\prime} }^2}=\frac{k_1k_2M_2\cos^2 \theta-k_2^2M_1\sin \theta\cos \theta }{k_1k_2M_2\sin^2 \theta+k_2^2M_1\sin \theta\cos \theta}.
\end{eqnarray}
After replacing $k_1k_2M_2$ with $(k_2^2-k_1^2)\tan 2\theta/2$ and simplification, we have:
\begin{eqnarray}
	    \frac{- m_{\hat{Z} }^2}{m_{\hat{Z}^{\prime} }^2}=\frac{k_2^2\tan ^2\theta-k_1^2}{k_2^2-k_1^2\tan ^2\theta}=\frac{k_2^2\sin ^2\theta-k_1^2\cos ^2\theta}{k_2^2\cos ^2\theta-k_1^2\sin ^2\theta}.
\end{eqnarray}
This means the amplitude is truly proportional to the momentum transfer, and we see that the DPDM model do possess the cancellation mechanism.

\section{Scalar induced kinetic mixing from $H$ and $\Phi$}
\label{app:scalarKM}

For Eq.~\eqref{Luv}, after electroweak and $\mathrm{U}(1)_\mathrm{Z^{\prime}}$ sysmmetry breaking, the neutral CP-even component of $H$ and $\Phi$ can be written as
\begin{eqnarray}
    H=\begin{pmatrix}
        G^{+}\\
	\frac{v_{H}+h_{H}+iG^{0}}{\sqrt{2} }
    \end{pmatrix}, \Phi=\frac{v_{\Phi}+h_{\Phi}}{\sqrt{2} },
\end{eqnarray}
and their mass terms take the form
\begin{eqnarray}
    -\frac{1}{2}\begin{pmatrix}
	h_{H}&h_{\Phi}
    \end{pmatrix}\begin{pmatrix}
	2\lambda_{H}v_{H}^2 & \lambda_{H\Phi}v_{H}v_{\Phi}\\
	\lambda_{H\Phi}v_{H}v_{\Phi} & 2\lambda_{\Phi}v_{\Phi}^2
    \end{pmatrix}\begin{pmatrix}
        h_{H}\\
	h_{\Phi}
    \end{pmatrix}=
    -\frac{1}{2}\begin{pmatrix}
	h_{H}&h_{\Phi}
    \end{pmatrix}\mathcal{M}_{H}^2\begin{pmatrix}
        h_{H}\\
	h_{\Phi}
    \end{pmatrix},
\end{eqnarray}
where $\lambda_{H}$ is the SM Higgs quartic coupling contained in $\mathcal{L}_{SM}$. The mass eigenstates $h_1,h_2$ are obtained via 
\begin{eqnarray}
    \begin{pmatrix}
        h_1\\
	h_2
    \end{pmatrix}=\begin{pmatrix}
    \cos \alpha & \sin \alpha\\
    -\sin \alpha & \cos \alpha
    \end{pmatrix}\begin{pmatrix}
        h_{H}\\
	h_{\Phi}
    \end{pmatrix}, \tan 2\alpha=\frac{\lambda_{H\Phi}v_{H}v_{\Phi}}{\lambda_{\Phi}v_{\Phi}^2-\lambda_{H}v_{H}^2}.
    \label{eq:Hmixangle}
\end{eqnarray}
In the gauge basis, $h_{H}$ couples only to the SM $Z$ boson while $h_{\Phi}$ couples only to $Z^{\prime}$. After rotating to $h_1$ and $h_2$, both eigenstates couple to $Z$ and $Z^{\prime}$, and their scalar loop contribution to the kinetic mixing in Eq.~\eqref{DPDM} can be obtained from Eq.~\eqref{kinmix} as
\begin{eqnarray}
    \epsilon_{\mathrm{scalar}}=\frac{1}{48\pi^2}(g_{Z}^{(1)}g_{Z^{\prime}}^{(1)}\ln \frac{\mu^2}{m_{h_{1}}^2}+g_{Z}^{(2)}g_{Z^{\prime}}^{(2)}\ln \frac{\mu^2}{m_{h_{2}}^2}),
    \label{eq:epsscalar}
\end{eqnarray}
where $g_{Z}^{(a)}$ and $g_{Z^{\prime}}^{(a)}$ are the $h_{a}ZZ$ and $h_{a}Z^{\prime}Z^{\prime}$ couplings ($a=1,2$ ) that follow from Eq.~\eqref{eq:Hmixangle}. For scalars in Eq.~\eqref{Luvt}, though $\Phi$ being a doublet, the mass terms of the two scalars and the following derivation of the $\epsilon_{\mathrm{scalar}}$ are the same as the above.   
\begin{acknowledgments}
    This work is supported in part by program for scientific research start-up funds of Guangdong Ocean University.
\end{acknowledgments}

\bibliography{ref}% ref.bib ..
\end{document}